\documentclass[journal,letterpaper]{IEEEtran}

\usepackage{cite}
\usepackage[pdftex]{graphicx}
\graphicspath{{./figures/}}
\DeclareGraphicsExtensions{.pdf}
\usepackage{amsmath}
\usepackage{algorithm}
\usepackage{algorithmic}

\usepackage[caption=false,font=footnotesize]{subfig}
\usepackage{url}
\usepackage[hidelinks]{hyperref}
\begin{document}
\title{Evaluation of Neuromorphic Spike Encoding of Sound Using Information Theory}

\author{Ahmad~El Ferdaoussi,
        Éric~Plourde,
        and~Jean~Rouat.\\[0.5cm]
        NECOTIS Research Lab, Université de Sherbrooke, QC, Canada.\\
        E-mail: \{Ahmad.El.Ferdaoussi, Jean.Rouat, Eric.Plourde\}@USherbrooke.ca
\thanks{This work was supported by the \textit{Fonds de Recherche du Québec -- Nature et Technologie.}}}

\maketitle

\begin{abstract}
The problem of spike encoding of sound consists in transforming a sound waveform into spikes. It is of interest in many domains, including the development of audio-based spiking neural networks, where it is the first and most crucial stage of processing. Many algorithms have been proposed to perform spike encoding of sound. However, a systematic approach to quantitatively evaluate their performance is currently lacking.
We propose the use of an information-theoretic framework to solve this problem. Specifically, we evaluate the coding efficiency of four spike encoding algorithms on two coding tasks that consist of coding the fundamental characteristics of sound: frequency and amplitude. The algorithms investigated are: Independent Spike Coding, Send-on-Delta coding, Ben's Spiker Algorithm, and Leaky Integrate-and-Fire coding.
Using the tools of information theory, we estimate the information that the spikes carry on relevant aspects of an input stimulus. We find disparities in the coding efficiencies of the algorithms, where Leaky Integrate-and-Fire coding performs best.
The information-theoretic analysis of their performance on these coding tasks provides insight on the encoding of richer and more complex sound stimuli.
\end{abstract}

\begin{IEEEkeywords}
Audio signal processing, neural coding, spike encoding, spiking neural networks, mutual information. 
\end{IEEEkeywords}

\section{Introduction}
\label{sect:introduction}
\IEEEPARstart{S}{ensory} systems are thought to efficiently encode their inputs \cite{Chalk2017}. It follows that models of such sensory systems ought to efficiently encode their inputs as well. The spike encoding of sound is an important processing step in models of the auditory system because any information lost at the spike encoding stage cannot be recovered at later processing stages. A multitude of algorithms can be used to convert sound to spikes. However, it is not clear if some encode their inputs more efficiently than others. To the best of our knowledge, the performance of spike encoding algorithms has only been evaluated through decoding, which refers either to performance evaluation on classification tasks or the ability to reconstruct the original stimulus from spikes. Decoding has important limitations, which we outline later, that limit the quantitative evaluation of spike encoding algorithms.

We propose the use of information theory to systematically evaluate the performance of spike encoding algorithms. We did not find any information-theoretic investigation of spike encoding algorithms for sound. However, the usage of mutual information here bears similarities to the one in \cite{Palmer2015} where spike encoding of stimuli is studied in the salamander's retina. In this work, we evaluate the performance of four widely used sound-to-spike encoding algorithms using the proposed approach by directly estimating the information that they capture on input sound stimuli.

We create two spike encoding tasks. In the first task, a sound stimulus is built from synthetic segments of frequency-modulated signals, and the goal is to estimate the information on the instantaneous frequency of the sound that is carried in the population encoded response. In the second task, a sound stimulus is built from synthetic segments of amplitude-modulated signals, and the goal is to estimate the information on the instantaneous amplitude of the sound that is encoded by a single encoding unit. The stimuli used in these two coding tasks, frequency and amplitude modulations, fundamentally characterize the acoustic waveform. As such, the performance of spike encoding algorithms on these coding tasks is relevant to the coding of richer and more complex sounds.

In the rest of this article, we provide background on the evaluation of spike encoding algorithms, and discuss decoding. We then present the methods related to the proposed approach based on information theory, as well as the results of the evaluation on the two spike encoding tasks. In the discussion, we give a synthesis of these results and consider the limitations of the study, followed by a conclusion.

\section{Evaluation of spike encoding algorithms}

The problem of converting sound into spikes, is of interest mainly in three application areas.
First, it is relevant in computational models of the peripheral auditory system that aim to reproduce properties of audition \cite{Meddis2010}.
Second, it is used in neuromorphic event-based microphones, known as silicon cochleae \cite{Liu2010a,Liu2014}, which output asynchronous spikes from analog sound.
Finally, it is fundamental in spiking neural networks (SNNs) that use spike representations as input.

The most widely used biologically plausible auditory features in these applications are spectro-temporal representations of sound called cochleagrams. These are obtained from cochlear filter banks and they roughly approximate the probabilities of auditory nerve fiber (ANF) spiking \cite{Patterson1992}. Filter banks avoid the block-based processing of Fast Fourier Transforms, which makes their outputs shift-invariant \cite{Pichevar2011}. They are the standard features used in SNNs for audio applications \cite{Verstraeten2006,Verstraeten2007,Schrauwen2008,Legenstein2008,Klampfl2013,Zhang2015,Jin2018,Loiselle2005}.

In the next two subsections, we first present the decoding approach and its limitations and then explore an information-theoretic approach to evaluating spike encoding algorithms as an alternative to the decoding approach.

\subsection{The decoding approach}
Spike encoding algorithms have been evaluated using the decoding approach. The term decoding can refer to either classification or stimulus reconstruction, which both consist in recovering (i.e. decoding) information about the original stimulus from the spikes. For example, classification is used in \cite{Verstraeten2005} to compare three auditory SNN front-ends, and stimulus reconstruction is used in \cite{Petro2020} to compare four spike encoding algorithms.

The decoding approach has several important limitations, the two most important being the information loss caused by decoding, and the influence of the choice of models used in the decoding results.

\paragraph{Information loss} The most important limitation of the decoding approach to studying the neural code is the information loss due to the further processing of the neuronal response that is necessary in the context of decoding. Thus, if one were to use decoding to compare different spike encoding methods, one would not be utilizing the total information available in the spikes but only a fraction of this total information, referred to as the residual information \cite{Quiroga2009,Quiroga2013}. This residual is only a lower bound on the total information captured by the spikes, whereas the total information can be estimated directly with information theory. In fact, a fundamental result in information theory, called the data processing inequality \cite{Cover2006}, says that one can never gain information with the further processing of data, only potentially lose it.

\paragraph{Model choice} One can never be assured of having the theoretically best decoder, only a lower bound on it. Therefore, the performance measured in decoding might say more about the choice of the functions introduced (e.g. classification model, signal decoding method) than the information captured by the spikes, as using different models or even different parameters can lead to different results \cite{Quiroga2013,Nelken2007}.
Moreover, if one were to use classification, one would have to pick a classification model, and this model might introduce parameters that need to be optimized.
This optimization can involve an overhead that could be limiting, depending on the choice of the model. Similarly, the use of stimulus reconstruction involves choosing a signal decoding method to convert the spikes back to a continuous time signal. This latter choice can be quite arbitrary, and in some cases very difficult.

\subsection{Proposed approach: information theory}
In view of these limitations to the decoding approach, we propose an information-theoretic framework to study spike encoding algorithms for sound. In addition to decoding, computational neuroscientists use information theory to study the neural code, which is a more rigorous approach \cite{Borst1999}. To the best of our knowledge, information theory has not yet been used to study artificial spike encoding algorithms.

In contrast to decoding, information theory measures the total information captured by the spikes without introducing models or additional processing that causes loss of information. It is model-free and multivariate, which makes it possible to study the information content with respect to different aspects of input stimuli. Furthermore, optimization based on mutual information maximization between the spikes and the stimulus is straightforward as a performance metric since mutual information curves tend to have concave shapes relative to the firing rate. This is because spike trains that are too dense or too sparse have low entropy, and thus carry little information.

We thus propose to use the mutual information between the stimuli and the encoded spikes as a quantitative metric to measure the information captured by a given encoding algorithm, which will be detailed in the next section.

\section{Methods}
\label{sect:methods}

In this section, we present in detail the stimulus generation protocol, the cochleagram extraction method, the four investigated spike encoding algorithms, and the procedure for estimating the mutual information and coding efficiency of the algorithms. 

\subsection{Stimulus generation}
\label{subsect:stimulus}
\begin{figure}[!t]
	\centering
	\subfloat[]{\includegraphics[width=\columnwidth]{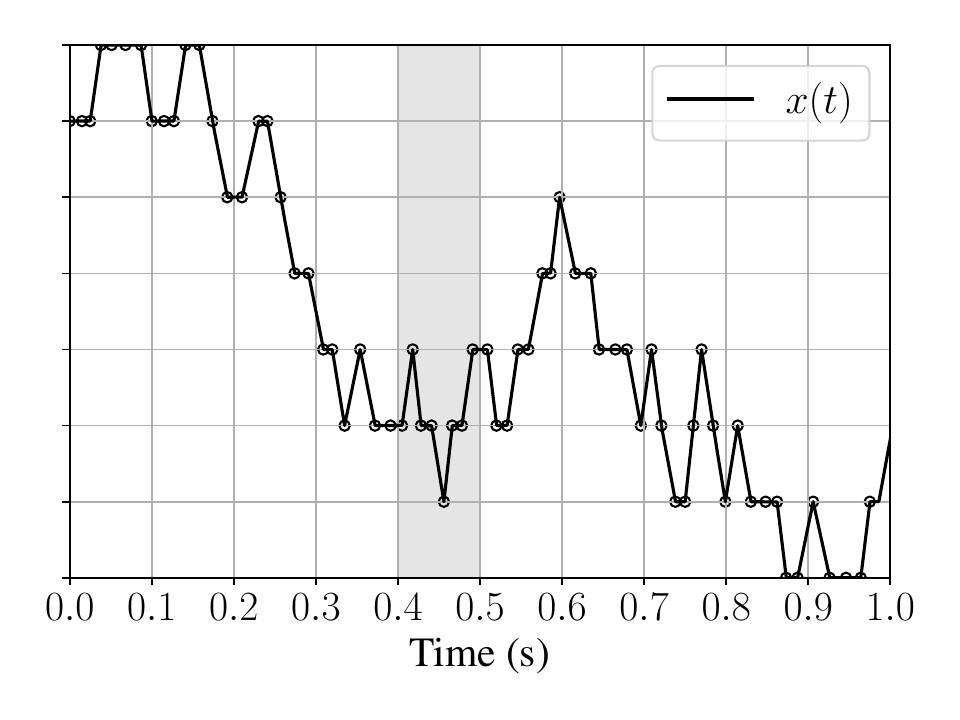}}

	\subfloat[]{\includegraphics[width=\columnwidth]{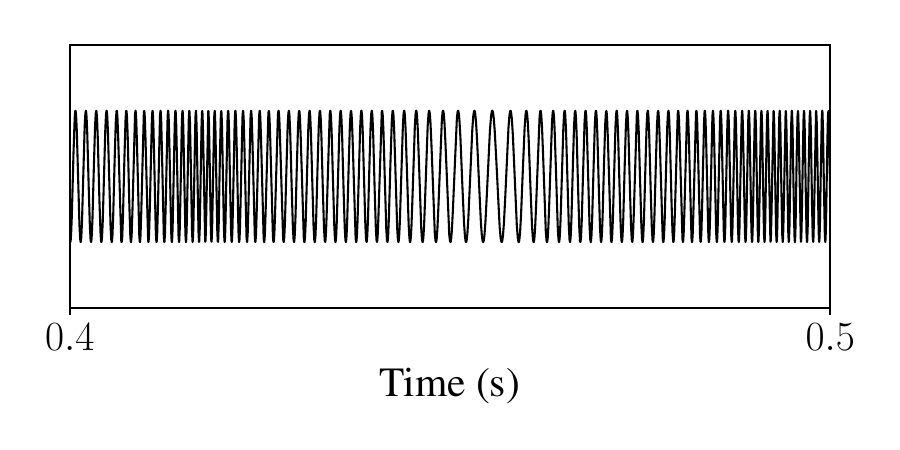}}
	
	\subfloat[]{\includegraphics[width=\columnwidth]{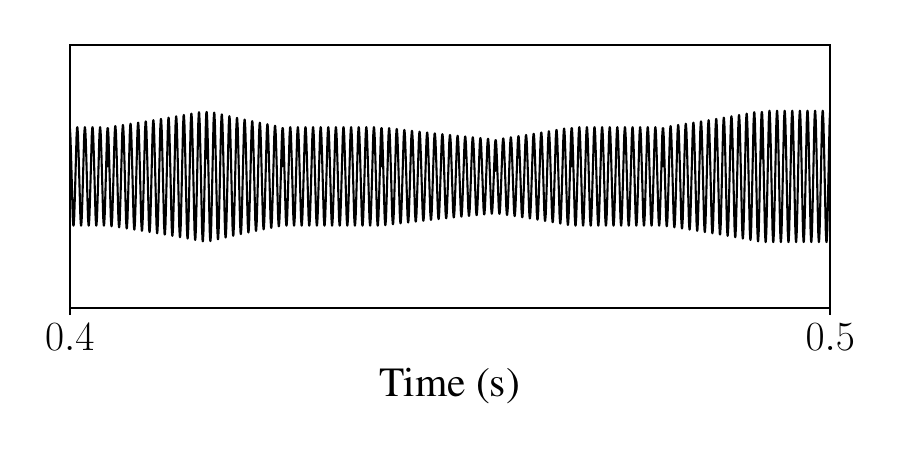}}
	
	\caption{(a) The first second of $x(t)$, a function that describes the variation in time of the sound characteristics, and from which the sound stimuli are built. A random walk is generated (circles in the graphic), and the vertices are joined to form linear pieces whose durations vary randomly. $x(t)$ is then mapped onto logarithmic scales of the sound characteristics, and modulated to form the sound stimulus for each characteristic. (b) and (c) The modulations of the sound waveforms for the 100 ms shaded duration in (a) for the frequency coding task (b) and the amplitude coding task (c).}
	\label{fig:characteristic}
\end{figure}

First, a long one-dimensional random walk is generated on an 8 point bounded set. From this random walk, a piecewise linear time function $x(t)$ is built by joining the successive random walk points into linear pieces. To introduce more randomness, the duration of the linear pieces is generated uniformly between 10 ms and 20 ms (Fig. \ref{fig:characteristic}a). The idea is to use $x(t)$ to build the sound stimulus such that $x(t)$ describes the value of the instantaneous sound characteristic as it varies in time (frequency or amplitude).

For the frequency coding task, the set of points is mapped onto a quasi-logarithmic frequency scale (ERB, equivalent rectangular bandwidth scale \cite{Moore1996}) ranging from 100 Hz to 10 kHz. This gives a function of instantaneous frequency that is used to build the stimulus of frequency modulations.

In the amplitude coding task, the set of points is mapped onto a logarithmic amplitude scale ranging from 0.1 to unity. This is then multiplied with a pure tone 1 kHz cosine to give an amplitude-modulated stimulus. A log scale is chosen to reflect the nonlinear perception of sound pressure in the cochlea.

Both stimuli last 300 seconds and are sampled at 32 kHz. The sound waveforms are illustrated in Fig. \ref{fig:characteristic} (b) and (c).

\subsection{Cochleagram}
\label{subsect:cochleagram}
Before encoding the sound stimuli into spikes, a spectral representation is extracted in the form of a cochleagram based on Gammatone filters \cite{Patterson1992}. This gives an approximation of auditory nerve fiber spike probabilities from the stimuli. The impulse response of a Gammatone filter simulates that of an ANF. The Gammatone filter is relatively simple and widely used to model the response of the basilar membrane \cite{Meddis2010}.

The filter's output approximates the vibrations of the basilar membrane. Next, inner hair cell effects are modeled by half-wave rectification of the filter's output, followed by cubic-root compression and low-pass filtering at 10 Hz to extract the envelope of the oscillations. Finally, the signal is decimated to 1 kHz, and normalized to a maximum of 1 to form the cochleagram. With the 1 kHz sampling rate, there is no need to consider refractoriness when encoding into spikes, as ANFs have a maximum firing rate of about 1 kHz \cite{Schnupp2013}. The cochleagram is implemented using the \texttt{brian2hears} Python library \cite{Fontaine2011}.

\begin{figure}[!t]
	\centering
	\includegraphics[width=\columnwidth]{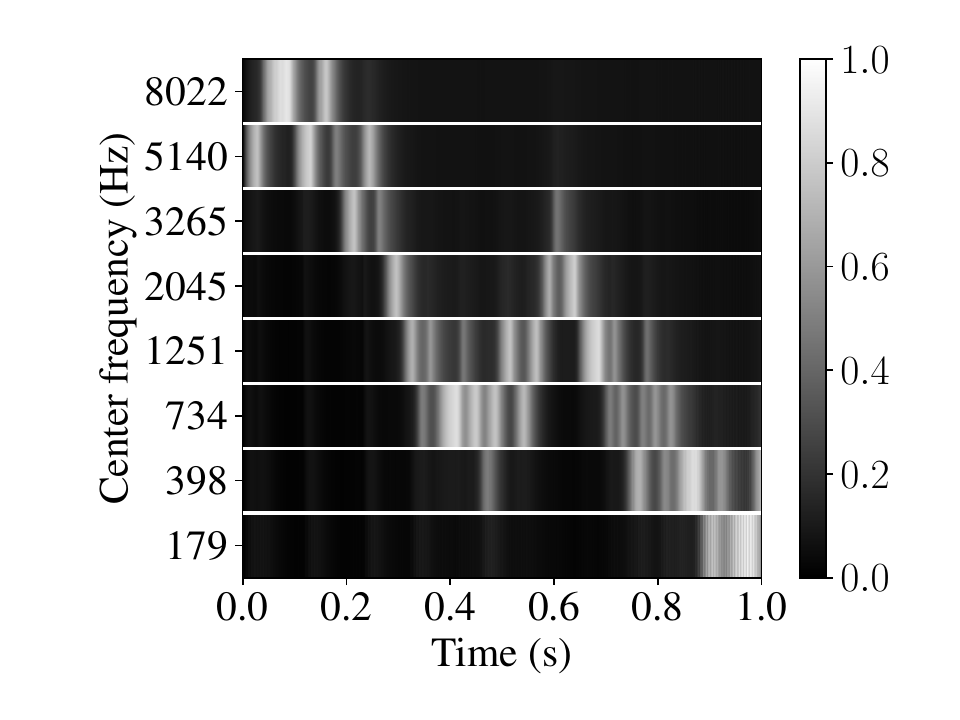}
	\caption{The first second of the cochleagram representation of the sound stimulus in the frequency coding task, using a parallel filter bank of 8 Gammatone filters with ERB-spaced center frequencies. The cochleagram models ANF spike probability.}
	\label{fig:cochleagram}
\end{figure}

For the frequency coding task, a bank of 8 Gammatone filters is used. The center frequencies of the filters are equally distributed on the ERB scale, in the frequency range of the sound stimulus (Fig. \ref{fig:cochleagram}). For the amplitude coding task, a single Gammatone filter is used with center frequency 1 kHz (corresponding to the carrier tone of the sound stimulus).

\subsection{Spike encoding}
\label{subsect:algorithms}
Four spike encoding algorithms are investigated. These algorithms encode the cochleagram's ANF spike probabilities, with each channel coded independently using a separate encoding unit. Note that we use ``encoding unit'' rather than ``neuron'' to avoid any confusion. Let $z(t)$ be the signal of a given channel. The pseudo-code for each of these algorithms is provided in the \hyperref[sect:appendix]{Appendix}.

\subsubsection{Independent Spike Coding (ISC)}
\label{sect:isc}
As mentioned previously, the cochleagram represents the instantaneous firing probability of ANFs. These probabilities can be used to emit spikes stochastically and independently, resulting in a non homogeneous Poisson process. Specifically, assuming the signal of a given cochleagram channel $z(t)$ is normalized between 0 and 1, the probability of a spike at time $t$ is
\begin{equation}
p(t) = z(t) \times a
\end{equation}
where $a$ is a scaling factor that controls the density of the spike train. A factor greater than unity will bias the spike train towards a higher firing rate, and vice versa. The scaling factor is the only parameter for this method.

ISC is used in \cite{Srinivasan2018} to encode cochleagrams for speech recognition. It is also the standard approach to generate spikes in computational models of the peripheral auditory system, such as those of Carney \cite{Carney1993} and Meddis \cite{Meddis1986}, where the probabilities are those of neurotransmitter release.
The Meddis model is used in \cite{Cramer2020} to produce the Heidelberg audio spike datasets for SNNs.

\subsubsection{Send-on-Delta (SoD) coding}
\label{sect:sod}
SoD \cite{Miskowicz2006} is basically the delta modulation method of analog-to-digital conversion, framed in a spike encoding context. An ON spike is emitted when the signal increases by a fixed value ($\Delta$), and an OFF spike is emitted when it decreases by the same value. This makes it possible to have a reconstruction of the original signal from the spikes, since between any two consecutive spikes, the signal has changed by $\Delta$, which is the only parameter to be optimized for this method. Specifically, the first and the $k$\textsuperscript{th} spikes are emitted at times
\begin{align}
t_0 &= \inf_{t \geq 0} \quad \{t, \quad |z(t) - z(0)| \geq \Delta\} \\
t_k &= \inf_{t \geq t_{k-1}} \{t, \quad |z(t) - z(t_{k-1})| \geq \Delta\}
\end{align}
with the spike being an ON event if the change is positive, and an OFF event if the change is negative. We use bipolar spikes to represent this ($+1$ and $-1$), although it is equivalent to using two different unipolar encoding units.

SoD is well known in the sensors community as an event-based sampling scheme.
It is used in \cite{Zimmer2019} for speech recognition, and is also used in the silicon cochlea model in \cite{Yang2016}. SoD has been discussed in other work under the name ``step-forward encoding'' \cite{Petro2020}.

\subsubsection{Ben's Spiker Algorithm (BSA)}
\label{sect:bsa}
BSA \cite{Schrauwen2003} is based on the reverse convolution of the signal with a linear finite impulse response (FIR) filter $h$ of length $M$. It emits spikes based on a simple heuristic comparing the filter's impulse response with the signal, such that the convolution of the resulting spike train with the filter gives a reconstruction of the original signal. Specifically, at each instant, two error metrics are computed:
\begin{align}
e_1 &= \sum_{k=0}^{M-1} |z(t-k) - h(M-1-k)|\\
e_2 &= \sum_{k=0}^{M-1} |z(t-k)|
\end{align}
If $e_1 \leq e_2 - \theta$, with $\theta$ a threshold, a spike is emitted and the filter is subtracted from the signal. This makes possible the reconstruction by convolving the spike train with the filter.

BSA was proposed as an improvement over previous similar stimulus estimation algorithms. It requires normalized signals, or alternatively to scale up the filter's coefficients, as the maximum signal value that can be reconstructed is equal to the sum of the coefficients.

Since BSA is a stimulus estimation method, the threshold $\theta$ is traditionally optimized by reconstruction of the stimulus, with the signal-to-noise ratio (SNR) as the error metric. However, a threshold that is optimal for stimulus reconstruction might not be optimal in information-theoretic terms. Therefore, the threshold is varied in this work to estimate information over a wide range of firing rates. BSA's parameters are the filter length $M$, the threshold $\theta$, and the filter's cutoff frequency. The cutoff frequency is set at 10 Hz, since the cochleagram is low-pass filtered at 10 Hz.

BSA has been used in works by Schrauwen and colleagues \cite{Verstraeten2005,Verstraeten2006,Verstraeten2007,Schrauwen2008}, Maass and colleagues \cite{Legenstein2008,Klampfl2013}, and Li and colleagues \cite{Zhang2015,Jin2018}.

\subsubsection{Leaky integrate-and-fire (LIF) coding}
\label{sect:lif}
In this method, the signal (i.e. a cochleagram channel) is fed directly as a membrane current to be integrated by a neuron with equation
\begin{equation}
\tau \frac{\mathrm{d}u}{\mathrm{d}t} = z(t) - u(t)
\label{eq:lif}
\end{equation}
where $u$ is the neuron's membrane potential, $\tau$ its time constant, and the signal to be encoded $z$ becomes the membrane current. We assume the resistance in the model to be equal to 1. If the potential crosses a threshold $u \geq \theta$, the neuron emits a spike, and its potential is reset to zero (resting potential).

One can pick alternative neuron models for similar encoding via current integration. The LIF model is chosen because it is the most widely used neuron model. LIF neurons can also be extended by including refractory periods, but they are not considered in this work to keep the number of parameters manageable. LIF's two parameters are then the neuron's time constant ($\tau$) and the firing threshold ($\theta$).

In \cite{Bellec2018} and \cite{Zenke2021}, LIF coding is used for speech recognition.
LIF coding is also the standard way to transduce the inner hair cell signal into spikes in silicon cochleae \cite{vanSchaik2010,Liu2010b}.

\subsection{Information theory measures}
\label{subsect:information}
The information theory measures, proposed here as quantitative metrics, are presented in this section. Mutual information is estimated between two discrete random variables that have to be defined from the stimulus and the spikes, respectively. This estimation is prone to statistical bias because of finite sampling, which is remedied by correcting the bias. From the mutual information, we derive the coding efficiency, which is the quantitative metric used to evaluate spike encoding algorithms in this work.

\subsubsection{Random variables}
\label{subsect:X}
To estimate the mutual information between the stimulus and the spikes, we need to define the corresponding discrete random variables.

The coding tasks consist of coding the sound characteristic (frequency or amplitude) of the stimulus. We recall that the variation in time of the relevant sound characteristic in each task is described by the time series $x(t)$. We create a discrete random variable $X$ with probability mass function $P(X=x)$ by quantizing $x(t)$ into 8 levels.

For the frequency coding task, the discrete random variable for the spikes $W$ is obtained from the vector of the population response in time of the 8 encoding units. For the amplitude coding task, $W$ is obtained from the vector of the time response of the encoding unit in the last 8 time steps.

\subsubsection{Mutual information and coding efficiency}

The mutual information $I(X;W;\Delta t)$ between the variable of the sound characteristic $X$ and the relevant spikes variable $W$ is estimated using \eqref{eq:mutual-information}, which is called the ``plug-in'' method~\cite{Palmer2015}
\begin{equation}
I(X;W;\Delta t) = \displaystyle\sum_{x\in X}\sum_{w\in W} P_{\Delta t}(x,w) \log_{2}\left( \frac{P_{\Delta t}(x,w)}{P(x) P(w)} \right)
\label{eq:mutual-information}
\end{equation}
where $P_{\Delta t}$ is the joint probability estimated with $x(t)$ shifted in time by discrete time steps $\Delta t$, and $P(x)$ and $P(w)$ the marginal probabilities. \eqref{eq:mutual-information} estimates in bits the information that the spikes carry concurrently on the sound characteristic ($\Delta t = 0$), on its past ($\Delta t < 0$), and on its future ($\Delta t > 0$).

In all mutual information estimations, the first 50 ms of the two time series is ignored to avoid onset effects of the cochleagram extraction.

We define the coding power (in bits) as the maximum of the mutual information curve for all time delays, which is the maximum amount of information, at every time instant, that $W$ encodes on $X$:
\begin{equation}
I_{\max} = \max_{\Delta t} I(X;W;\Delta t).
\end{equation}

The maximum coding power achievable is determined by the statistical structure of $X$ and is given by the entropy $H(X)$.
In other words, if $W$ completely characterizes $X$, then the mutual information is equal to $H(X)$, which thus forms a tight upper bound on the coding power and can only be attained by a perfect encoding. From this, we can define a measure of coding efficiency $\epsilon$ as the ratio of the coding power to the entropy of $X$,
\begin{equation}
\epsilon = \frac{I_\mathrm{max}}{H(X)}
\end{equation}
with the entropy of $X$ being
\begin{equation}
H(X) = - \displaystyle\sum_{x\in X} P(x) \log_{2} P(x).
\label{eq:entropy}
\end{equation}

The coding efficiency $\epsilon$ provides the performance metric for the evaluation of the spike encoding algorithms in this work, as changing the parameters of an encoding algorithm or choosing a different algorithm is only going to change the coding power, $I_\mathrm{max}$, while $H(X)$ will only depend on the stimulus.

\subsubsection{Sampling bias}
Finite sampling can make mutual information heavily biased (upwards). This is because finite sampling makes the reliable estimation of the joint probability mass function between the stimuli and the neural responses more difficult for more complex data. This is why we only use 8 encoding units.

To control for this bias, for each mutual information estimation, the spike variable $W$ is shuffled in time and the mutual information is re-estimated using this shuffled variable. This yields an ``error'' term, because this mutual information is theoretically null since the variables are independent after shuffling. The error term, normalized by the entropy of $X$, was less than 1.6\% for SoD, and less than 0.16\% for the other algorithms in the simulations. The error for SoD is an order of magnitude greater than for the other methods because SoD has bipolar spikes (ON and OFF).

With such low ``error'' terms, we can safely apply a bias correction method. Several approaches have been proposed for this \cite{Panzeri2007}. We use the quadratic extrapolation method \cite{Strong1998,Treves1995}.

\section{Task no. 1: Frequency coding}
\label{sect:results}

\subsection{The frequency coding task}
As mentioned previously, the sound stimulus is built from piece-wise segments of continuously-varying frequency modulations. Let $X$ be the discrete random variable of the frequency, built from $x(t)$. An eight-channel cochleagram is then extracted from the sound, and each channel is encoded independently by the algorithms into a spike matrix $\mathbf{W}$ of 8 rows. The objective for the algorithms is to encode the instantaneous frequency of the sound stimulus.

The frequency coding task is facilitated by the multichannel processing of the cochlear filter bank, and the instantaneous frequency of the sound stimulus will appear as a high spike probability in the cochleagram's corresponding channel.

\begin{figure}[!t]
	\centering
	\includegraphics[width=\columnwidth]{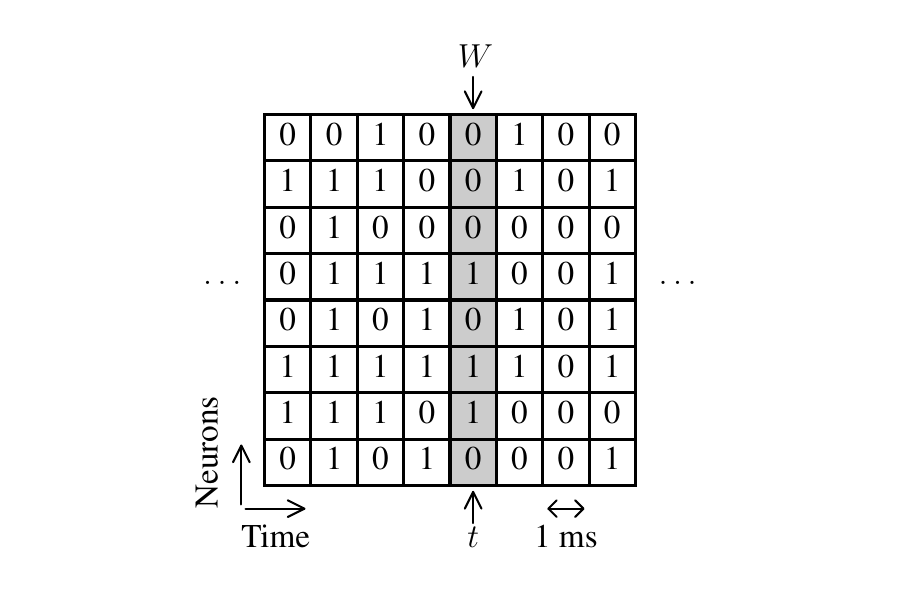}
	\caption{Spike variable $W$ in the frequency coding task, taken as the instantaneous population response in time of the encoding units. This presupposes an underlying population coding of frequency, consistent with the place coding of frequency in the cochlea.}
	\label{fig:spikes-frequency}
\end{figure}

We must extract from the spike matrix $\mathbf{W}$ some variable $W$, that presupposes an underlying neural code. In the cochlea, hair cells encode sound frequency via a place code. This is the basis of the tonotopic organization of the cochlea \cite{Schnupp2013}. Consistent with the place coding of frequency in the cochlea, we assume that the frequency information is encoded with a population code (i.e. in the instantaneous population response of the encoding units). By looking at this population code through time, we should be able to recover information on the sound frequency. Therefore, we build the spike variable $W$ as the instantaneous population response of the 8 encoding units. This variable can be seen as a list of the encoding units that fired at a given time instant (Fig. \ref{fig:spikes-frequency}).

\subsection{Mutual information}
\begin{figure}[!t]
	\centering
	\includegraphics[width=\columnwidth]{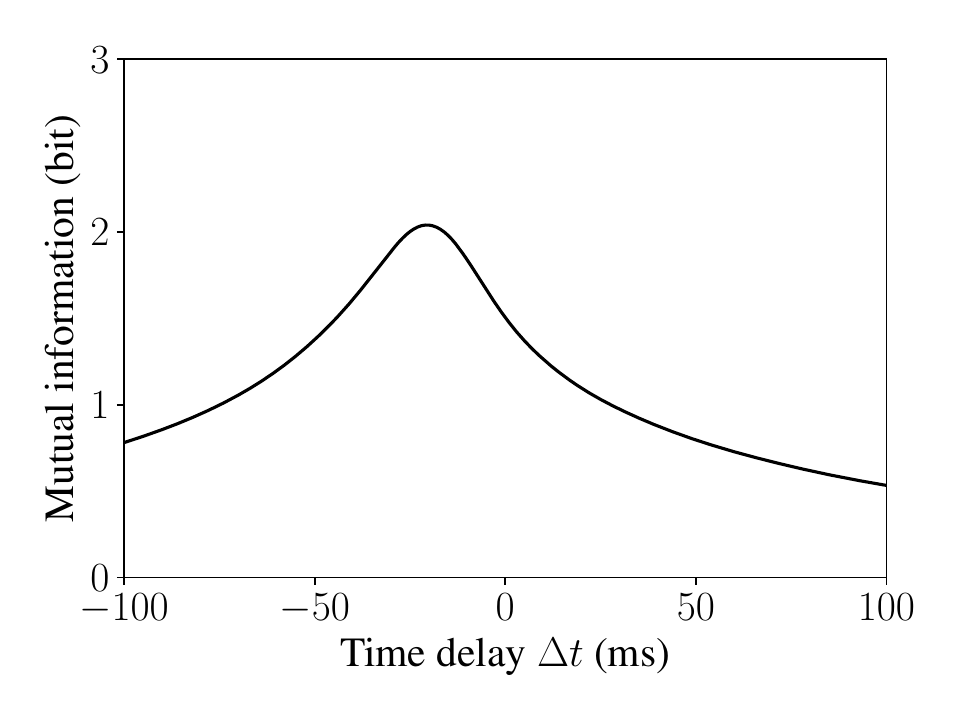}
	\caption{A typical example of a time-delayed mutual information curve. The maximum is in the past (for a negative time delay $\Delta t$). This means that at any given instant $t$, on average, the spike variable $W$ carries the maximum amount of information on the sound variable $X$ at $t' \leq t$, some time in the past. This maximum is the coding power.}
	\label{fig:mutual-information}
\end{figure}

The mutual information between the sound frequency $X$ and the spike variable $W$ can be estimated as $I(X;W)$ (Sect. \ref{subsect:information}). This, however, would only give the information that $W$ gives \textit{concurrently} on $X$, that is to say the information carried by $w$ at instant $t$ on $X$ at the same time instant. The processing involved in spike encoding can entail some latency, which means that $W$ at time $t$ can be maximally informative on $X$ some time in the past at $t' \leq t$. For this reason, the mutual information has to be estimated with different time delays between the two variables, that is $I(X;W;\Delta t)$, where $X$ is shifted in time by $\Delta t$ (Fig. \ref{fig:mutual-information}).

The maximum amount of information that the spikes variable $W$ carries on the sound frequency $X$ is the coding power $I_\mathrm{max}$, which is the maximum of the time-delayed mutual information curve. The maximum coding power achievable is $H(X)$, the entropy of $X$. The coding efficiency $\epsilon$ is the ratio of the coding power to the entropy of $X$ (Sect. \ref{subsect:information}).

\subsection{Results}
Contrary to a real spike train (for which time is continuous), an artificial spike train is discrete in time, and therefore has a fixed maximum firing rate. At the boundaries, where the spike train is too sparse or too dense, the unit is not capable of encoding much information. Therefore, mutual information curves tend to have concave shapes relative to firing rate.
We can define a spike density measure $\rho$ as a normalized mean firing rate comprised between 0 and 1. The spike density can be obtained by taking the mean over the whole absolute spike matrix $\overline{|\mathbf{W}|}$, with the absolute value accounting for the special case of bipolar spike trains. The spike density is then
\begin{equation}
\rho = \overline{|\mathbf{W}|}
\label{eq:spike-density}
\end{equation}
with zero being a completely silent spike train/matrix, and unity a completely saturated one.

\begin{figure}[!t]
	\centering
	\includegraphics[width=\columnwidth]{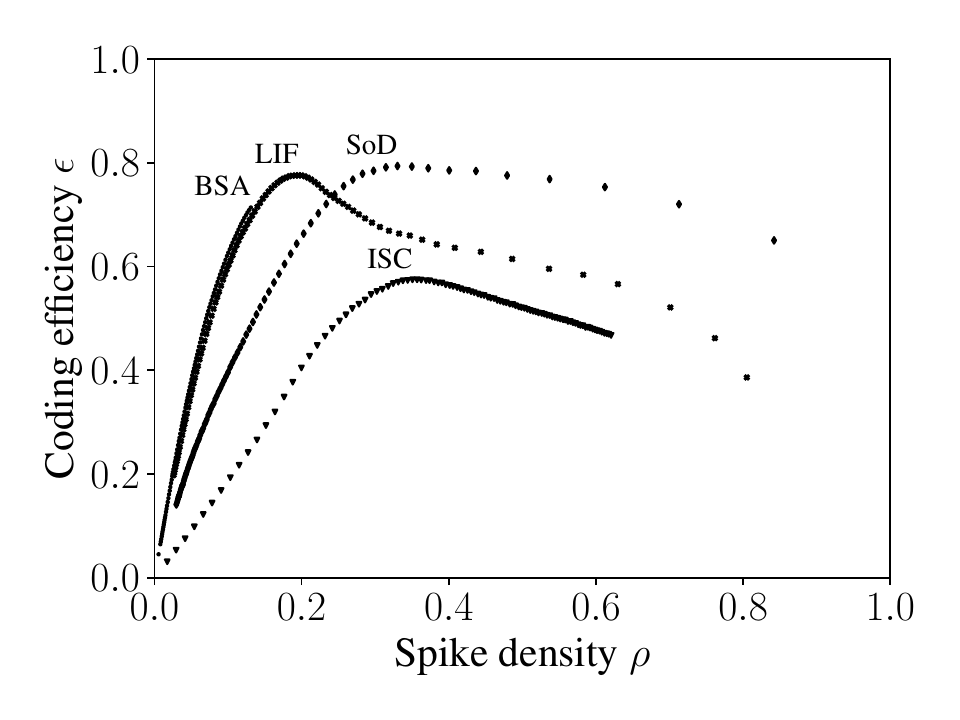}	
	\caption{Coding efficiency vs. spike density in the frequency coding task. The curves show how each algorithm's efficiency varies as a function of the spike density of the underlying spike trains. The curves are found by varying the parameters of each algorithm. For BSA, we only show the curve for the best filter length found ($M = 3$). Similarly, for LIF, we only show the curve for the best time constant found (the limit $\tau \to 0$).
	Standard errors of the means are under 0.002 for both axes over 5 trials, so the error bars are not shown. By design, BSA is limited in spike density, and achieves a maximum coding efficiency of $\sim 71\%$ at a spike density of $\sim 13\%$. LIF reaches the best coding efficiency ($\sim 80\%$) at a low spike density ($\sim 18\%$). SoD achieves a comparable coding efficiency but with a denser spike train ($\sim 33\%$).}
	\label{fig:results-frequency}
\end{figure}

We can then estimate the coding efficiency $\epsilon$ for different parameter configurations and plot it as a function of the unidimensional spike density scale. This shows how much of the information is captured by the spikes at different firing rates, and allows us to compare between the different algorithms. By finding the maximum of these curves, we find the best coding efficiency that the encoding algorithms can achieve on the coding task, as well as the spike density corresponding to it. The results on the frequency coding task are shown in Fig. \ref{fig:results-frequency}.

The stochastic nature of ISC introduces noise that degrades the spike train's information content, which can explain its low coding efficiency, as well as its requirement to spike a lot more than needed so as to maximize efficiency.

The good performance of SoD is not surprising since a frequency change in the stimulus occasions an onset and an offset in the cochleagram channels that causes jumps in the signals that SoD codes very well. Since SoD codes both onsets and offsets, this explains its higher spike density around its maximum coding efficiency, compared to LIF. However, SoD is the only algorithm that cannot sustain firing for a stable signal value (i.e. stable frequency), which is about a third of the duration of the stimulus. Therefore, there is no way of decoding a stable instantaneous frequency value over time.

BSA's coding efficiency reaches its maximum at the boundary of its threshold parameter. BSA is by design limited in the spike density range it produces, as its threshold cannot yield densities outside the range shown in the BSA curve in Fig. \ref{fig:results-frequency}.

Finally, LIF achieves a coding efficiency comparable to SoD but with fewer spikes.

\section{Task no. 2: Amplitude coding}
\subsection{The amplitude coding task}
The amplitude coding task is analogous to the previous one. The sound stimulus is built from piece-wise segments of continuously-varying amplitude modulations of a 1 kHz pure tone carrier frequency. $X$ is now the discrete random variable of the amplitude, built from $x(t)$. By contrast, only a single channel cochleagram is now used. Its center frequency is equal to that of the carrier tone of the sound. This channel is then coded by the algorithms into a single spike train $\mathbf{W}$, which is now a vector. The objective for the algorithms is to encode the instantaneous amplitude of the sound, which is represented as changes in the spike probability in the channel.

\begin{figure}[!t]
	\centering
	\includegraphics[width=\columnwidth]{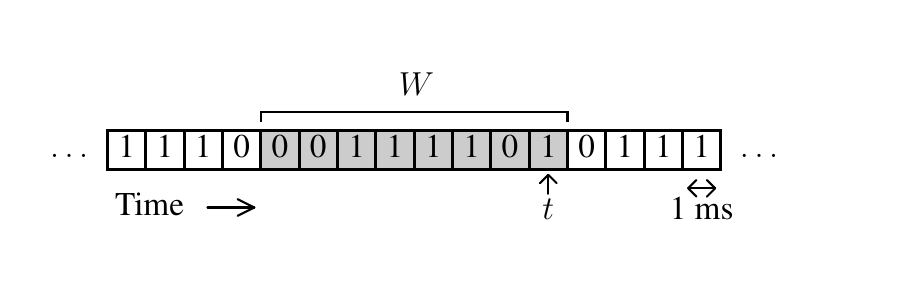}
	\caption{Spike variable $W$ from the spike train $\mathbf{W}$ in the amplitude coding task. $W$ is the spiking pattern over the last 8 time steps of the single encoding unit, consistent with the rate coding of the amplitude in the cochlea.}
	\label{fig:spikes-amplitude}
\end{figure}

We must also extract from the spike train $\mathbf{W}$ a variable $W$. In this case, we are dealing with a single spike train, where information is encoded in the timing of the spikes. In the cochlea, ANFs encode sound intensity with a rate code \cite{Schnupp2013}. Therefore, by looking at the spike pattern in a time window in the past, we should be able to recover information on the sound amplitude at a given instant in the past. We build the spike variable $W$ as the pattern of spikes over the last 8 time steps, which is a reasonably long enough time window to encode the amplitude of the stimulus. This variable can be seen as a list of spike times in the recent past (Fig. \ref{fig:spikes-amplitude}). Having defined $W$, the analysis is identical to the frequency coding task.

\subsection{Results}
\begin{figure}[!t]
	\centering
	\includegraphics[width=\columnwidth]{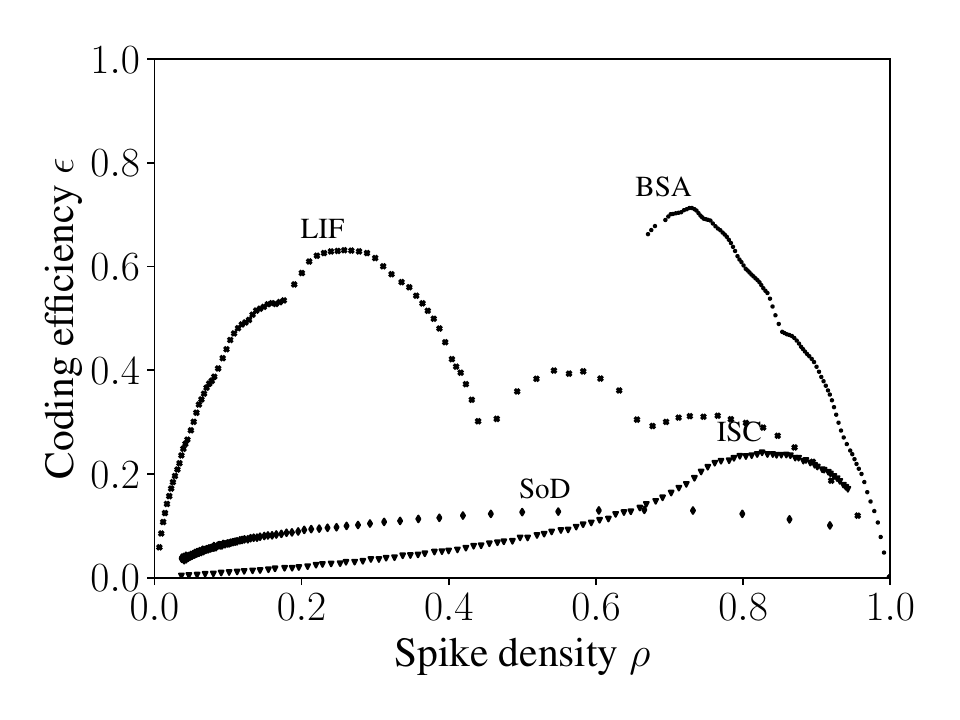}
	\caption{Coding efficiency vs. spike density in the amplitude coding task. The curves show how each algorithm's efficiency varies as a function of spike density of the underlying spike trains. For BSA, we only show the curve for the best filter length found ($M = 9$). Similarly, for LIF, we only show the curve for the best time constant found (the limit $\tau = 2$ ms).
	Standard errors of the means are under 0.002 for both axes over 5 trials, and therefore the error bars are not shown. ISC and SoD reach very low coding efficiencies. BSA achieves the highest coding efficiency at $\sim 71\%$ but at a saturated spike train of density $\sim 73\%$. LIF reaches a coding efficiency of $\sim 63\%$ at the lowest spike density $\sim 26\%$.}
	\label{fig:results-amplitude}
\end{figure}

As for the frequency coding task, the coding efficiency is plotted as a function of spike density. The results on the amplitude coding task are shown in Fig. \ref{fig:results-amplitude}.

The stochasticity of ISC again introduces noise that degrades the spike train's information content. This can explain its low coding efficiency peak at a very high spike density.

SoD is shown to have very poor coding efficiency, which is not surprising, since SoD encodes the changes in a signal. Intuitively, this means that the spike train can only inform on changes in levels (relative amplitude), not on the values of the levels (exact value of the amplitude). To recover the latter, one would need to know the initial level, and keep track of all the changes up to the present. For this reason, SoD might show good performance if it is the derivative of the amplitude that is the aspect of interest in the task at hand, rather than the exact amplitude value as in the task considered here.

BSA achieves the highest coding efficiency $\sim 71\%$, but at a saturated spike train of density $\sim 73\%$. BSA is by design limited in the spike density range that it can yield. Therefore, the curve shown in Fig. \ref{fig:results-amplitude} shows the whole spike density range of BSA on this task.

LIF, on the other hand, achieves $\sim 63\%$ coding efficiency at the lowest spike density $\sim 26\%$.

\section{Discussion and limitations}
There are many algorithms that can be used in the artificial spike encoding of sound. To the best of our knowledge, these algorithms have never been investigated for the information content of the spike trains they produce (i.e. using the information theory approach). They have only been studied in terms of their performance in tasks that further process the spike trains destructively, thereby causing information loss (i.e. using the decoding approach). In this work, four different spike encoding algorithms are compared in terms of their coding efficiency with respect to the two fundamental characteristics of sound: frequency and amplitude. The results show that the performance of these algorithms varies quite a bit.

\begin{itemize}
	\item \textbf{Independent Spike Coding}, which is the standard method in computational auditory modeling, is shown to have poor coding efficiency, which is not a surprise due to the stochastic nature of its spike generation, which introduces noise in the spike trains that degrades the information content.
	\item \textbf{Send-on-Delta coding}, which is an event-based sampling scheme that is popular in neuromorphic sensors, is shown to be unable to effectively code the amplitude of sound, as it codes only relative changes in amplitude, and not absolute changes. It should thus be avoided if the absolute sound level of the stimulus to be coded is an important aspect in the spike encoding problem.
	\item \textbf{Ben's Spiker Algorithm}, which is based on stimulus estimation, is found to perform well in both coding tasks. However, it tends to produce saturated spike trains due to its requirement to optimally reconstruct the stimulus. Moreover, BSA produces narrow spike density ranges and this does not leave much room for tuning the firing rate.
	\item \textbf{Leaky Integrate-and-Fire coding} shows the best performance in the spike encoding tasks considered here. The LIF mechanism is by design suited for the place coding of frequency and the rate coding of amplitude. The two parameters (time constant and firing threshold) offer much room to encode at a wide range of firing rates, and there is a trade-off to be found between them in terms of maximizing the encoded information.
\end{itemize}

This study has certain limitations. Firstly, a very specific spectro-temporal representation is used in this work: the cochleagram, which represents the average firing probabilities of ANFs. The performance of a spike encoding algorithm is ultimately contingent on the nature of the features to be encoded, and alternatives that could potentially represent the frequency-amplitude information in a fundamentally different way could lead to different results. However, the cochleagram is a widely used biologically plausible representation and is thus the candidate of choice for this study. Secondly, ANFs also have several important properties that we did not seek to reproduce in this work, like adaptation and phase-locking. 
Finally, it should not be forgotten that the sensory world in this work is very simple, consisting of collections of amplitude and frequency modulations, and that the stimuli that usually have to be converted into spikes are more rich and complex. However, the acoustic waveform is ultimately characterized by frequency and amplitude modulations, and therefore the performance of spike encoding algorithms on the two coding tasks presented here is relevant for more complex input.

\section{Conclusions}
We propose the use of information theory to evaluate the performance of spike encoding algorithms. To the best of our knowledge, this work is the first information-theoretic study of artificial spike encoding. We quantitatively evaluate the performance of four spike encoding algorithms of sound on two coding tasks. These tasks consist in coding the fundamental characteristics of the acoustic waveform: frequency and amplitude modulations. We find that coding efficiency differs for the investigated algorithms, and that they peak in coding efficiency at different spike densities. The best spike encoding algorithm depends on the nature of the sound stimulus to be converted into spikes. Moreover, coding efficiency strongly depends on the firing rate of the underlying encoding units (spike density). In this work, LIF coding appears to be the most general for the coding tasks considered here.

\appendix[Pseudo-code for spike encoding algorithms]
\label{sect:appendix}

Algorithms \ref{algo:isc} through \ref{algo:lif} give the pseudo-code for the four investigated spike encoding algorithms in this work. Each algorithm's input is a signal to be encoded (i.e. a cochleagram channel signal $z(t)$) and a parameter configuration. The output is the encoded spike train $w$.

\begin{algorithm}[H]
	\caption{Independent Spike Coding (Sect. \ref{sect:isc})}
	\label{algo:isc}
	\begin{algorithmic}[1]
		\REQUIRE normalized signal $z$, scaling factor $a$
		\ENSURE spike train $w$
		\STATE $w \leftarrow \mathrm{zeros}(\mathrm{length}(z))$
		\FOR{$t = 1:\mathrm{length}(z)$}
		\IF{$\mathrm{rand}\_\mathrm{uniform}(0,1) < z(t) \times a$}
		\STATE $w(t) \leftarrow 1$
		\ENDIF
		\ENDFOR
		\RETURN $w$
	\end{algorithmic}
\end{algorithm}

\begin{algorithm}[H]
	\caption{Send-on-Delta coding (Sect. \ref{sect:sod})}
	\label{algo:sod}
	\begin{algorithmic}[1]
		\REQUIRE signal $z$, delta $\Delta$
		\ENSURE bipolar spike train $w$
		\STATE $w \leftarrow \mathrm{zeros}(\mathrm{length}(z))$
		\STATE $b \leftarrow z(1)$
		\FOR{$t = 2:\mathrm{length}(z)$}
		\STATE $d \leftarrow z(t) - b$
		\IF{$d > \Delta$}
		\STATE $w(t) \leftarrow 1$
		\STATE $b \leftarrow b + \Delta$
		\ENDIF
		\IF{$d < - \Delta$}
		\STATE $w(t) \leftarrow -1$
		\STATE $b \leftarrow b - \Delta$
		\ENDIF
		\ENDFOR
		\RETURN $w$
	\end{algorithmic}
\end{algorithm}

\begin{algorithm}[H]
	\caption{Ben's Spiker Algorithm (Sect. \ref{sect:bsa})}
	\label{algo:bsa}
	\begin{algorithmic}[1]
		\REQUIRE normalized signal $z$, filter $h$, threshold $\theta$
		\ENSURE spike train $w$
		\STATE $M \leftarrow \mathrm{length}(h)$
		\STATE $w \leftarrow \mathrm{zeros}(\mathrm{length}(z))$
		\FOR{$t = M:\mathrm{length}(z)$}
		\STATE $e_1 \leftarrow 0$
		\STATE $e_2 \leftarrow 0$
		\FOR{$k = 0:M-1$}
		\STATE $e_1 \leftarrow e_1+|z(t-k)-h(M-k)|$
		\STATE $e_2 \leftarrow e_2+|z(t-k)|$
		\ENDFOR
		\IF{$e_1 \leq e_2 - \theta$}
		\STATE $w(t) \leftarrow 1$
		\FOR{$k = 0:M-1$}
		\STATE $z(t-k) \leftarrow z(t-k) - h(M-k)$
		\ENDFOR
		\ENDIF
		\ENDFOR
		\RETURN $w$
	\end{algorithmic}
\end{algorithm}

\begin{algorithm}[H]
	\caption{Leaky Integrate-and-Fire coding (Sect. \ref{sect:lif})}
	\label{algo:lif}
	\begin{algorithmic}[1]
		\REQUIRE signal $z$, time constant $\tau$, threshold $\theta$
		\ENSURE spike train $w$
		\STATE $w \leftarrow \mathrm{zeros}(\mathrm{length}(z))$
		\STATE $u \leftarrow \mathrm{zeros}(1+\mathrm{length}(z))$
		\STATE $u(1) \leftarrow z(1)$
		\FOR{$t = 1:\mathrm{length}(z)$}
		\STATE $u(t+1) \leftarrow u(t) \, \mathrm{e}^{-\frac{1}{\tau}} + z(t)$
		\IF{$u(t+1) \geq \theta$}
		\STATE $w(t) \leftarrow 1$
		\STATE $u(t+1) \leftarrow 0$
		\ENDIF
		\ENDFOR
		\RETURN $w$
	\end{algorithmic}
\end{algorithm}

\bibliographystyle{IEEEtran}
\bibliography{IEEEabrv,refs}

\begin{thebibliography}{10}
\providecommand{\url}[1]{#1}
\csname url@samestyle\endcsname
\providecommand{\newblock}{\relax}
\providecommand{\bibinfo}[2]{#2}
\providecommand{\BIBentrySTDinterwordspacing}{\spaceskip=0pt\relax}
\providecommand{\BIBentryALTinterwordstretchfactor}{4}
\providecommand{\BIBentryALTinterwordspacing}{\spaceskip=\fontdimen2\font plus
\BIBentryALTinterwordstretchfactor\fontdimen3\font minus
  \fontdimen4\font\relax}
\providecommand{\BIBforeignlanguage}[2]{{%
\expandafter\ifx\csname l@#1\endcsname\relax
\typeout{** WARNING: IEEEtran.bst: No hyphenation pattern has been}%
\typeout{** loaded for the language `#1'. Using the pattern for}%
\typeout{** the default language instead.}%
\else
\language=\csname l@#1\endcsname
\fi
#2}}
\providecommand{\BIBdecl}{\relax}
\BIBdecl

\bibitem{Chalk2017}
\BIBentryALTinterwordspacing
M.~Chalk, O.~Marre, and G.~Tka{\v{c}}ik, ``Toward a unified theory of
  efficient, predictive, and sparse coding,'' \emph{Proceedings of the National
  Academy of Sciences}, vol. 115, no.~1, pp. 186--191, Dec. 2017. [Online].
  Available: \url{https://doi.org/10.1073/pnas.1711114115}
\BIBentrySTDinterwordspacing

\bibitem{Palmer2015}
\BIBentryALTinterwordspacing
S.~E. Palmer, O.~Marre, M.~J. Berry, and W.~Bialek, ``Predictive information in
  a sensory population,'' \emph{Proceedings of the National Academy of
  Sciences}, vol. 112, no.~22, pp. 6908--6913, May 2015. [Online]. Available:
  \url{https://doi.org/10.1073/pnas.1506855112}
\BIBentrySTDinterwordspacing

\bibitem{Meddis2010}
\BIBentryALTinterwordspacing
R.~Meddis and E.~A. Lopez-Poveda, ``Auditory periphery: From pinna to auditory
  nerve,'' in \emph{Computational Models of the Auditory System}, R.~Meddis,
  E.~Lopez-Poveda, A.~Popper, and R.~Fay, Eds.\hskip 1em plus 0.5em minus
  0.4em\relax Springer {US}, 2010, pp. 7--38. [Online]. Available:
  \url{https://doi.org/10.1007/978-1-4419-5934-8_2}
\BIBentrySTDinterwordspacing

\bibitem{Liu2010a}
\BIBentryALTinterwordspacing
S.-C. Liu and T.~Delbruck, ``Neuromorphic sensory systems,'' \emph{Current
  Opinion in Neurobiology}, vol.~20, no.~3, pp. 288--295, Jun. 2010. [Online].
  Available: \url{https://doi.org/10.1016/j.conb.2010.03.007}
\BIBentrySTDinterwordspacing

\bibitem{Liu2014}
\BIBentryALTinterwordspacing
S.-C. Liu, T.~Delbruck, G.~Indiveri, A.~Whatley, and R.~Douglas, ``Silicon
  cochleas,'' in \emph{Event-Based Neuromorphic Systems}.\hskip 1em plus 0.5em
  minus 0.4em\relax John Wiley {\&} Sons, Ltd, Dec. 2014, pp. 71--90. [Online].
  Available: \url{https://doi.org/10.1002/9781118927601.ch4}
\BIBentrySTDinterwordspacing

\bibitem{Patterson1992}
\BIBentryALTinterwordspacing
R.~Patterson, K.~Robinson, J.~Holdsworth, D.~McKeown, C.~Zhang, and
  M.~Allerhand, ``Complex sounds and auditory images,'' in \emph{Auditory
  Physiology and Perception}, Y.~Cazals, K.~Horner, and L.~Demany, Eds.\hskip
  1em plus 0.5em minus 0.4em\relax Pergamon, 1992, pp. 429--446. [Online].
  Available: \url{https://doi.org/10.1016/b978-0-08-041847-6.50054-x}
\BIBentrySTDinterwordspacing

\bibitem{Pichevar2011}
\BIBentryALTinterwordspacing
R.~Pichevar, H.~Najaf-Zadeh, L.~Thibault, and H.~Lahdili, ``Auditory-inspired
  sparse representation of audio signals,'' \emph{Speech Communication},
  vol.~53, no.~5, pp. 643--657, May 2011. [Online]. Available:
  \url{https://doi.org/10.1016/j.specom.2010.09.008}
\BIBentrySTDinterwordspacing

\bibitem{Verstraeten2006}
\BIBentryALTinterwordspacing
D.~Verstraeten, B.~Schrauwen, and D.~Stroobandt, ``Reservoir-based techniques
  for speech recognition,'' in \emph{Proceedings of the 2006 {IEEE}
  International Joint Conference on Neural Networks}.\hskip 1em plus 0.5em
  minus 0.4em\relax Institute of Electrical and Electronics Engineers ({IEEE}),
  Jul. 2006, pp. 1050--1053. [Online]. Available:
  \url{https://doi.org/10.1109/ijcnn.2006.246804}
\BIBentrySTDinterwordspacing

\bibitem{Verstraeten2007}
\BIBentryALTinterwordspacing
D.~Verstraeten, B.~Schrauwen, M.~D'Haene, and D.~Stroobandt, ``An experimental
  unification of reservoir computing methods,'' \emph{Neural Networks},
  vol.~20, no.~3, pp. 391--403, Apr. 2007. [Online]. Available:
  \url{https://doi.org/10.1016/j.neunet.2007.04.003}
\BIBentrySTDinterwordspacing

\bibitem{Schrauwen2008}
\BIBentryALTinterwordspacing
B.~Schrauwen, M.~D'Haene, D.~Verstraeten, and J.~V. Campenhout, ``Compact
  hardware liquid state machines on {FPGA} for real-time speech recognition,''
  \emph{Neural Networks}, vol.~21, no. 2--3, pp. 511--523, Mar. 2008. [Online].
  Available: \url{https://doi.org/10.1016/j.neunet.2007.12.009}
\BIBentrySTDinterwordspacing

\bibitem{Legenstein2008}
\BIBentryALTinterwordspacing
R.~Legenstein, D.~Pecevski, and W.~Maass, ``A learning theory for
  reward-modulated spike-timing-dependent plasticity with application to
  biofeedback,'' \emph{{PLoS} Computational Biology}, vol.~4, no.~10, p.
  e1000180, Oct. 2008. [Online]. Available:
  \url{https://doi.org/10.1371/journal.pcbi.1000180}
\BIBentrySTDinterwordspacing

\bibitem{Klampfl2013}
\BIBentryALTinterwordspacing
S.~Klampfl and W.~Maass, ``Emergence of dynamic memory traces in cortical
  microcircuit models through {STDP},'' \emph{Journal of Neuroscience},
  vol.~33, no.~28, pp. 11\,515--11\,529, Jul. 2013. [Online]. Available:
  \url{https://doi.org/10.1523/jneurosci.5044-12.2013}
\BIBentrySTDinterwordspacing

\bibitem{Zhang2015}
\BIBentryALTinterwordspacing
Y.~Zhang, P.~Li, Y.~Jin, and Y.~Choe, ``A digital liquid state machine with
  biologically inspired learning and its application to speech recognition,''
  \emph{{IEEE} Transactions on Neural Networks and Learning Systems}, vol.~26,
  no.~11, pp. 2635--2649, Nov. 2015. [Online]. Available:
  \url{https://doi.org/10.1109/tnnls.2015.2388544}
\BIBentrySTDinterwordspacing

\bibitem{Jin2018}
\BIBentryALTinterwordspacing
Y.~Jin, W.~Zhang, and P.~Li, ``Hybrid macro/micro level backpropagation for
  training deep spiking neural networks,'' in \emph{Advances in Neural
  Information Processing Systems}, S.~Bengio, H.~Wallach, H.~Larochelle,
  K.~Grauman, N.~Cesa-Bianchi, and R.~Garnett, Eds.\hskip 1em plus 0.5em minus
  0.4em\relax Curran Associates, Inc., 2018, pp. 7005--7015. [Online].
  Available:
  \url{https://papers.nips.cc/paper/2018/file/3fb04953d95a94367bb133f862402bce-Paper.pdf}
\BIBentrySTDinterwordspacing

\bibitem{Loiselle2005}
\BIBentryALTinterwordspacing
S.~Loiselle, J.~Rouat, D.~Pressnitzer, and S.~Thorpe, ``Exploration of rank
  order coding with spiking neural networks for speech recognition,'' in
  \emph{Proceedings of the 2005 {IEEE} International Joint Conference on Neural
  Networks}.\hskip 1em plus 0.5em minus 0.4em\relax Institute of Electrical and
  Electronics Engineers ({IEEE}), 2005, pp. 2076--2080. [Online]. Available:
  \url{https://doi.org/10.1109/ijcnn.2005.1556220}
\BIBentrySTDinterwordspacing

\bibitem{Verstraeten2005}
\BIBentryALTinterwordspacing
D.~Verstraeten, B.~Schrauwen, D.~Stroobandt, and J.~V. Campenhout, ``Isolated
  word recognition with the liquid state machine: a case study,''
  \emph{Information Processing Letters}, vol.~95, no.~6, pp. 521--528, Sep.
  2005. [Online]. Available: \url{https://doi.org/10.1016/j.ipl.2005.05.019}
\BIBentrySTDinterwordspacing

\bibitem{Petro2020}
\BIBentryALTinterwordspacing
B.~Petro, N.~Kasabov, and R.~M. Kiss, ``Selection and optimization of temporal
  spike encoding methods for spiking neural networks,'' \emph{{IEEE}
  Transactions on Neural Networks and Learning Systems}, vol.~31, no.~2, pp.
  358--370, Feb. 2020. [Online]. Available:
  \url{https://doi.org/10.1109/tnnls.2019.2906158}
\BIBentrySTDinterwordspacing

\bibitem{Quiroga2009}
\BIBentryALTinterwordspacing
R.~Q. Quiroga and S.~Panzeri, ``Extracting information from neuronal
  populations: information theory and decoding approaches,'' \emph{Nature
  Reviews Neuroscience}, vol.~10, no.~3, pp. 173--185, Mar. 2009. [Online].
  Available: \url{https://doi.org/10.1038/nrn2578}
\BIBentrySTDinterwordspacing

\bibitem{Quiroga2013}
\BIBentryALTinterwordspacing
------, ``Decoding and information theory in neuroscience,'' in
  \emph{Principles of Neural Coding}, R.~Q. Quiroga and S.~Panzeri, Eds.\hskip
  1em plus 0.5em minus 0.4em\relax {CRC} Press, May 2013, pp. 156--179.
  [Online]. Available: \url{https://doi.org/10.1201/b14756-12}
\BIBentrySTDinterwordspacing

\bibitem{Cover2006}
\BIBentryALTinterwordspacing
T.~M. Cover and J.~A. Thomas, \emph{Elements of Information Theory},
  2nd~ed.\hskip 1em plus 0.5em minus 0.4em\relax Wiley, 2006. [Online].
  Available: \url{https://doi.org/10.1002/047174882x}
\BIBentrySTDinterwordspacing

\bibitem{Nelken2007}
\BIBentryALTinterwordspacing
I.~Nelken and G.~Chechik, ``Information theory in auditory research,''
  \emph{Hearing Research}, vol. 229, no. 1--2, pp. 94--105, Jul. 2007.
  [Online]. Available: \url{https://doi.org/10.1016/j.heares.2007.01.012}
\BIBentrySTDinterwordspacing

\bibitem{Borst1999}
\BIBentryALTinterwordspacing
A.~Borst and F.~E. Theunissen, ``Information theory and neural coding,''
  \emph{Nature Neuroscience}, vol.~2, no.~11, pp. 947--957, Nov. 1999.
  [Online]. Available: \url{https://doi.org/10.1038/14731}
\BIBentrySTDinterwordspacing

\bibitem{Moore1996}
\BIBentryALTinterwordspacing
B.~C.~J. Moore and B.~R. Glasberg, ``A revision of zwicker's loudness model,''
  \emph{Acta Acustica united with Acustica}, vol.~82, no.~2, pp. 335--345,
  1996. [Online]. Available:
  \url{https://www.ingentaconnect.com/content/dav/aaua/1996/00000082/00000002/art00020}
\BIBentrySTDinterwordspacing

\bibitem{Schnupp2013}
\BIBentryALTinterwordspacing
J.~Schnupp, ``Coding in the auditory system,'' in \emph{Principles of Neural
  Coding}, R.~Q. Quiroga and S.~Panzeri, Eds.\hskip 1em plus 0.5em minus
  0.4em\relax {CRC} Press, May 2013, pp. 196--221. [Online]. Available:
  \url{https://doi.org/10.1201/b14756-14}
\BIBentrySTDinterwordspacing

\bibitem{Fontaine2011}
\BIBentryALTinterwordspacing
B.~Fontaine, D.~F.~M. Goodman, V.~Benichoux, and R.~Brette, ``Brian hears:
  Online auditory processing using vectorization over channels,''
  \emph{Frontiers in Neuroinformatics}, vol.~5, 2011. [Online]. Available:
  \url{https://doi.org/10.3389/fninf.2011.00009}
\BIBentrySTDinterwordspacing

\bibitem{Srinivasan2018}
\BIBentryALTinterwordspacing
G.~Srinivasan, P.~Panda, and K.~Roy, ``{SpiLinC}: Spiking liquid-ensemble
  computing for unsupervised speech and image recognition,'' \emph{Frontiers in
  Neuroscience}, vol.~12, Aug. 2018. [Online]. Available:
  \url{https://doi.org/10.3389/fnins.2018.00524}
\BIBentrySTDinterwordspacing

\bibitem{Carney1993}
\BIBentryALTinterwordspacing
L.~H. Carney, ``A model for the responses of low-frequency auditory-nerve
  fibers in cat,'' \emph{The Journal of the Acoustical Society of America},
  vol.~93, no.~1, pp. 401--417, Jan. 1993. [Online]. Available:
  \url{https://doi.org/10.1121/1.405620}
\BIBentrySTDinterwordspacing

\bibitem{Meddis1986}
\BIBentryALTinterwordspacing
R.~Meddis, ``Simulation of mechanical to neural transduction in the auditory
  receptor,'' \emph{The Journal of the Acoustical Society of America}, vol.~79,
  no.~3, pp. 702--711, Mar. 1986. [Online]. Available:
  \url{https://doi.org/10.1121/1.393460}
\BIBentrySTDinterwordspacing

\bibitem{Cramer2020}
\BIBentryALTinterwordspacing
B.~Cramer, Y.~Stradmann, J.~Schemmel, and F.~Zenke, ``The heidelberg spiking
  data sets for the systematic evaluation of spiking neural networks,''
  \emph{{IEEE} Transactions on Neural Networks and Learning Systems}, pp.
  1--14, 2020. [Online]. Available:
  \url{https://doi.org/10.1109/tnnls.2020.3044364}
\BIBentrySTDinterwordspacing

\bibitem{Miskowicz2006}
\BIBentryALTinterwordspacing
M.~Miskowicz, ``Send-on-delta concept: An event-based data reporting
  strategy,'' \emph{Sensors}, vol.~6, no.~1, pp. 49--63, Jan. 2006. [Online].
  Available: \url{https://doi.org/10.3390/s6010049}
\BIBentrySTDinterwordspacing

\bibitem{Zimmer2019}
\BIBentryALTinterwordspacing
R.~Zimmer, T.~Pellegrini, S.~F. Singh, and T.~Masquelier, ``Technical report:
  supervised training of convolutional spiking neural networks with pytorch,''
  Tech. Rep., 2019. [Online]. Available: \url{https://arxiv.org/abs/1911.10124}
\BIBentrySTDinterwordspacing

\bibitem{Yang2016}
\BIBentryALTinterwordspacing
M.~Yang, C.-H. Chien, T.~Delbruck, and S.-C. Liu, ``A 0.5v 55{$\mu$}w
  64{$\times$}2-channel binaural silicon cochlea for event-driven stereo-audio
  sensing,'' \emph{{IEEE} Journal of Solid-State Circuits}, vol.~51, no.~11,
  pp. 2554--2569, Nov. 2016. [Online]. Available:
  \url{https://doi.org/10.1109/jssc.2016.2604285}
\BIBentrySTDinterwordspacing

\bibitem{Schrauwen2003}
B.~{Schrauwen} and J.~{Van Campenhout}, ``Bsa, a fast and accurate spike train
  encoding scheme,'' in \emph{Proceedings of the 2003 {IEEE} International
  Joint Conference on Neural Networks}, vol.~4.\hskip 1em plus 0.5em minus
  0.4em\relax Institute of Electrical and Electronics Engineers ({IEEE}), 2003,
  pp. 2825--2830.

\bibitem{Bellec2018}
\BIBentryALTinterwordspacing
G.~Bellec, D.~Salaj, A.~Subramoney, and R.~L.~W. Maass, ``Long short-term
  memory and learning-to-learn in networks of spiking neurons,'' in
  \emph{Advances in Neural Information Processing Systems}, S.~Bengio,
  H.~Wallach, H.~Larochelle, K.~Grauman, N.~Cesa-Bianchi, and R.~Garnett,
  Eds.\hskip 1em plus 0.5em minus 0.4em\relax Curran Associates, Inc., 2018,
  pp. 787--797. [Online]. Available:
  \url{https://proceedings.neurips.cc/paper/2018/file/c203d8a151612acf12457e4d67635a95-Paper.pdf}
\BIBentrySTDinterwordspacing

\bibitem{Zenke2021}
\BIBentryALTinterwordspacing
F.~Zenke and T.~P. Vogels, ``The remarkable robustness of surrogate gradient
  learning for instilling complex function in spiking neural networks,''
  \emph{Neural Computation}, vol.~33, no.~4, pp. 899--925, 2021. [Online].
  Available: \url{https://doi.org/10.1162/neco_a_01367}
\BIBentrySTDinterwordspacing

\bibitem{vanSchaik2010}
\BIBentryALTinterwordspacing
A.~van Schaik, T.~J. Hamilton, and C.~Jin, ``Silicon models of the auditory
  pathway,'' in \emph{Computational Models of the Auditory System}, A.~P. R.~F.
  R.~Meddis, E.A. Lopez-Poveda, Ed.\hskip 1em plus 0.5em minus 0.4em\relax
  Springer {US}, 2010, pp. 261--276. [Online]. Available:
  \url{https://doi.org/10.1007/978-1-4419-5934-8_10}
\BIBentrySTDinterwordspacing

\bibitem{Liu2010b}
\BIBentryALTinterwordspacing
S.-C. Liu, A.~van Schaik, B.~A. Mincti, and T.~Delbruck, ``Event-based
  64-channel binaural silicon cochlea with {Q} enhancement mechanisms,'' in
  \emph{Proceedings of the 2010 {IEEE} International Symposium on Circuits and
  Systems}.\hskip 1em plus 0.5em minus 0.4em\relax Institute of Electrical and
  Electronics Engineers ({IEEE}), May 2010. [Online]. Available:
  \url{https://doi.org/10.1109/iscas.2010.5537164}
\BIBentrySTDinterwordspacing

\bibitem{Panzeri2007}
\BIBentryALTinterwordspacing
S.~Panzeri, R.~Senatore, M.~A. Montemurro, and R.~S. Petersen, ``Correcting for
  the sampling bias problem in spike train information measures,''
  \emph{Journal of Neurophysiology}, vol.~98, no.~3, pp. 1064--1072, Sep. 2007.
  [Online]. Available: \url{https://doi.org/10.1152/jn.00559.2007}
\BIBentrySTDinterwordspacing

\bibitem{Strong1998}
\BIBentryALTinterwordspacing
S.~P. Strong, R.~Koberle, R.~R. de~Ruyter~van Steveninck, and W.~Bialek,
  ``Entropy and information in neural spike trains,'' \emph{Physical Review
  Letters}, vol.~80, no.~1, pp. 197--200, Jan. 1998. [Online]. Available:
  \url{https://doi.org/10.1103/physrevlett.80.197}
\BIBentrySTDinterwordspacing

\bibitem{Treves1995}
\BIBentryALTinterwordspacing
A.~Treves and S.~Panzeri, ``The upward bias in measures of information derived
  from limited data samples,'' \emph{Neural Computation}, vol.~7, no.~2, pp.
  399--407, Mar. 1995. [Online]. Available:
  \url{https://doi.org/10.1162/neco.1995.7.2.399}
\BIBentrySTDinterwordspacing

\end{thebibliography}

\end{document}